\documentclass[reprint,amsmath,amssymb,superscriptaddress,prl,floatfix,nobibnotes]{revtex4-1}
\usepackage{graphicx}
\usepackage{bm}

\usepackage[compact]{titlesec}

\titlespacing\section{0pt}{4pt plus 2pt minus 2pt}{0pt plus 2pt minus 2pt}
\titlespacing\subsection{0pt}{4pt plus 2pt minus 2pt}{0pt plus 2pt minus 2pt}
\titlespacing\subsubsection{0pt}{4pt plus 2pt minus 2pt}{0pt plus 2pt minus 2pt}

\begin{document}
\title{Impact of Quenched Oxygen Disorder on Charge Density Wave Order in YBa$_2$Cu$_3$O$_{6+x}$} 

\author{A. J. Achkar}
\affiliation{Department of Physics and Astronomy, University of Waterloo, Waterloo, N2L 3G1, Canada}
\author{X. Mao}
\affiliation{Department of Physics and Astronomy, University of Waterloo, Waterloo, N2L 3G1, Canada}
\author{Christopher McMahon}
\affiliation{Department of Physics and Astronomy, University of Waterloo, Waterloo, N2L 3G1, Canada}
\author{R. Sutarto}
\affiliation{Canadian Light Source, Saskatoon, Saskatchewan, S7N 2V3, Canada}
\author{F. He}
\affiliation{Canadian Light Source, Saskatoon, Saskatchewan, S7N 2V3, Canada}
\author{Ruixing Liang}
\affiliation{Department of Physics and Astronomy, University of British Columbia, Vancouver,V6T 1Z1, Canada}
\author{D. A. Bonn}
\affiliation{Department of Physics and Astronomy, University of British Columbia, Vancouver,V6T 1Z1, Canada}
\author{W. N. Hardy}
\affiliation{Department of Physics and Astronomy, University of British Columbia, Vancouver,V6T 1Z1, Canada}
\author{D. G. Hawthorn}
\affiliation{Department of Physics and Astronomy, University of Waterloo, Waterloo, N2L 3G1, Canada} 

\begin{abstract}
The competition between superconductivity and charge density wave (CDW) order in underdoped cuprates has now been widely reported, but the role of disorder in this competition has yet to be fully resolved. A central question is whether disorder sets the length scale of the CDW order, for instance by pinning charge density fluctuations or disrupting an otherwise long range order. Using resonant soft x-ray scattering, we investigate the sensitivity of CDW order in YBa$_2$Cu$_3$O$_{6+x}$ (YBCO) to varying levels of oxygen disorder. We find that quench cooling YBCO$_{6.67}$ (YBCO$_{6.75}$) crystals to destroy their o-V and o-VIII (o-III) chains decreases the intensity of the CDW superlattice peak by a factor of 1.9 (1.3), but has little effect on the CDW correlation length, incommensurability, and temperature dependence. This reveals that while quenched oxygen disorder influences the CDW order parameter, the spatial extent of the CDW order is insensitive to the level of quenched oxygen disorder and may instead be a consequence of competition with superconductivity. 
\end{abstract}


\date{\today}

\maketitle

Charge density wave (CDW) order has been solidified as a generic property and principal competitor to superconductivity (SC) in underdoped cuprate superconductors through its observation in YBa$_2$Cu$_3$O$_{6+x}$ (YBCO),\cite{Ghiringhelli12,Chang12,Achkar12,Blackburn13,Blanco-Canosa13,Wu11,LeBoeuf13} Bi-based, \cite{Hanaguri04,Kohsaka07,Fujita12,daSilvaNeto14,Comin14a}, La-based,\cite{Wu12} and Hg-based cuprates.\cite{Tabis14} Despite important differences in crystal structure and levels of disorder in these cuprates,\cite{Eisaki04} the spatial extent of CDW order is relatively short range in all cases. The origin of this common short range character is not currently understood. A widely held view is that disorder plays the role of either pinning charge density fluctuations or disrupting an otherwise long-range order.\cite{Robertson06,DelMaestro06,Nie13} These possibilities have been used to describe the effect of disorder and impurities in the cuprates. For example, apical oxygen vacancies in Bi$_{2+y}$Sr$_{2-y}$CaCu$_2$O$_{8+x}$ were argued to pin a CDW checkerboard state.\cite{Zeljkovic2012} The substitution of spinless Zn atoms for Cu atoms in YBCO was argued to disrupt [enhance] CDW [spin density wave (SDW)] correlations in the vicinity of the Zn defects.\cite{Blanco-Canosa13} Within such interpretations, the common short range character of CDW order in the cuprates is associated with each material's specific defect properties and crystal structure. However, it is also possible that such descriptions only apply due to the high defect strength, masking a more generic and intrinsic origin of this length scale, such as the competition of CDW order with superconductivity.\cite{Sachdev04,Hayward13}

To address this question, we turn to high-purity, oxygen ordered YBCO. With regards to defects, YBCO represents a special case in the cuprates since stoichiometric, ultra-high purity crystals can be grown with low levels of cation disorder.\cite{Liang98} Doping of the CuO$_2$ planes occurs by the addition of oxygen atoms into the chain layer, which can organize into a number of ortho-ordered phases depending on the oxygen content.\cite{Uimin94,Schleger95,Andersen99} Disorder in these CuO chains have been shown by microwave conductivity studies of quasiparticle scattering in YBCO$_{6.5}$ to be the dominant source of weak-limit scattering,\cite{Bobowski10} indicating that the most influential defects in YBCO reside in the chain layer.\cite{Eisaki04} As previously established, the oxygen ordered states can be intentionally destroyed by heating YBCO crystals to modest temperatures and subsequently quench cooling to prevent the formation of chain order.\cite{Andersen99,Liang00,Zimmermann03,Bobowski10}  This allows for individual crystals of YBCO to be investigated with varying degrees of disorder.

In this Letter, we exploit this means of controlling disorder in YBCO to study the effect of quenched disorder on CDW order in the cuprates using resonant soft x-ray scattering (RSXS). Our main finding is that disordering the chains decreases the CDW scattering intensity, but has little impact on the CDW correlation length ($\xi^\text{CDW}$), incommensurability or $T$ dependence. This reveals that while disorder influences the CDW order parameter, the length scale of the CDW order is insensitive to the level of disorder. We argue that this observation is difficult to reconcile with simple pictures of disorder-induced pinning or of order nucleating around defects and suggest that the short range character of the CDW order has an intrinsic origin, possibly rooted in the competition between CDW order and superconductivity. We also discuss how the disorder effect studied here can be contrasted to disorder effects in other cuprates. 

RSXS and x-ray absorption spectroscopy (XAS) measurements were performed at the Canadian Light Source's REIXS beamline \cite{Hawthorn11a} using high purity single crystals of YBCO with o-V ($T_c\!=\!64.5$ K, $p\!=\!0.116$, $x\!=\!0.667$), o-VIII ($T_c\!=\!65.5$ K, $p\!=\!0.118$, $x\!=\!0.667$) and o-III ($T_c\!=\!75.2$ K, $p\!=\!0.133$, $x\!=\!0.75$) oxygen ordering.\cite{Liang98,Liang06}  The samples were oriented with the $ac$ plane parallel to the scattering plane (with the $c$-axis normal to the sample surface). The orientation was confirmed using $(0\ 0\ 2)$ and $(\pm1\ 0\ 2)$ Bragg reflections at 2 keV. Scattering was performed with $\sigma$ polarized light and XAS was measured by total fluorescence yield (TFY).  

\begin{figure}
\centering
\resizebox{\columnwidth}{!}{\includegraphics{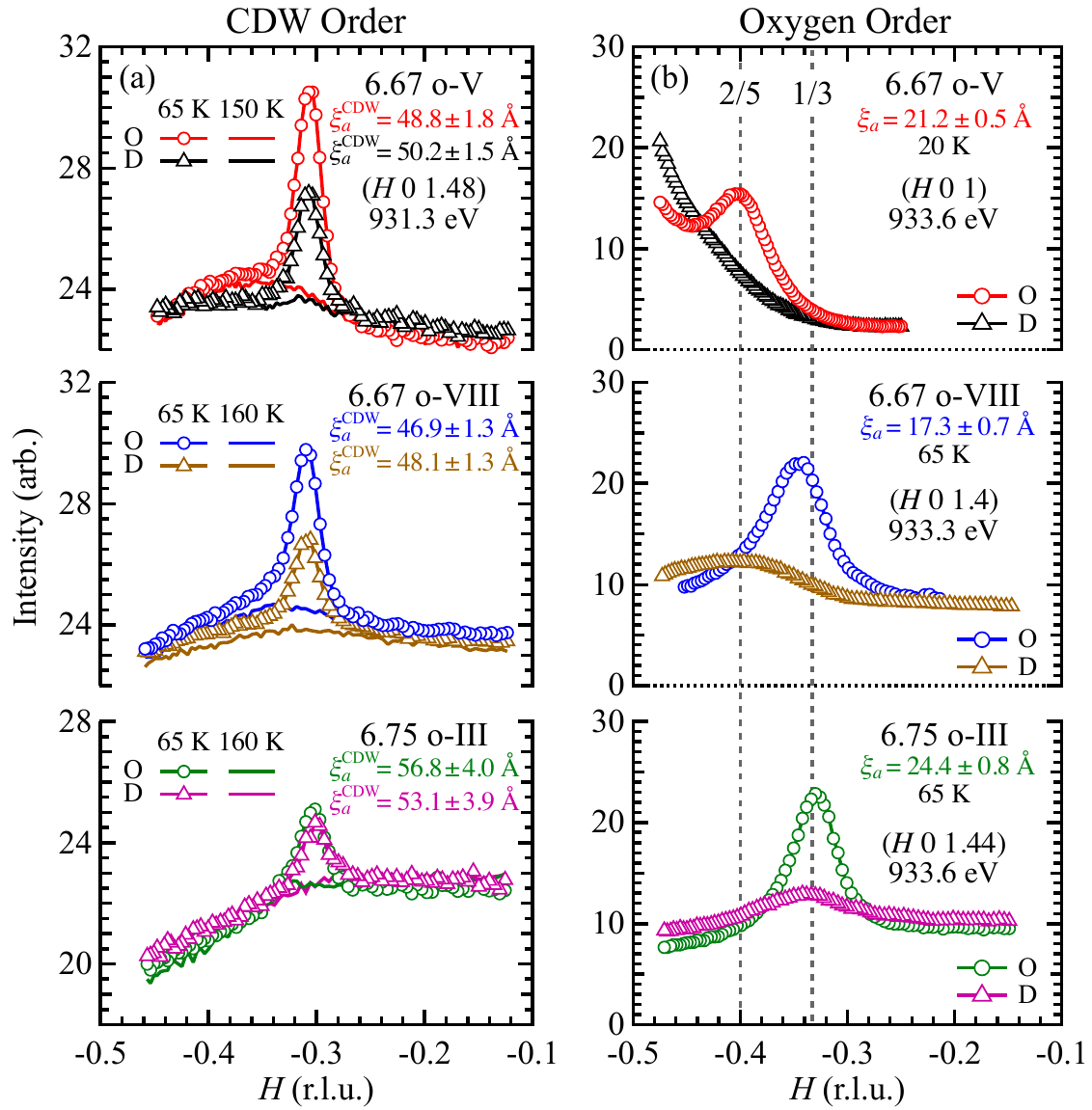}}
\caption{($H$ 0 $L$) scans through the (a) CDW peaks and (b) oxygen chain ordering superstructure peaks in o-V, o-VIII and o-III ordered YBCO before (O) and after (D) heating the samples to disorder the oxygen in the chain layer. In panel (a) a small contribution of the oxygen order superstructure reflection is still visible for the oxygen ordered states at the photon energy 931.3 eV, where CDW order is most pronounced. The correlation lengths $\xi^\text{CDW}_a$ given in (a) are for the ordered and disordered states of the crystals and the ortho structure correlation lengths $\xi_a$ in (b) are for the ordered state.}
\label{fig1}
\end{figure}

\begin{figure}
\centering
\resizebox{\columnwidth}{!}{\includegraphics{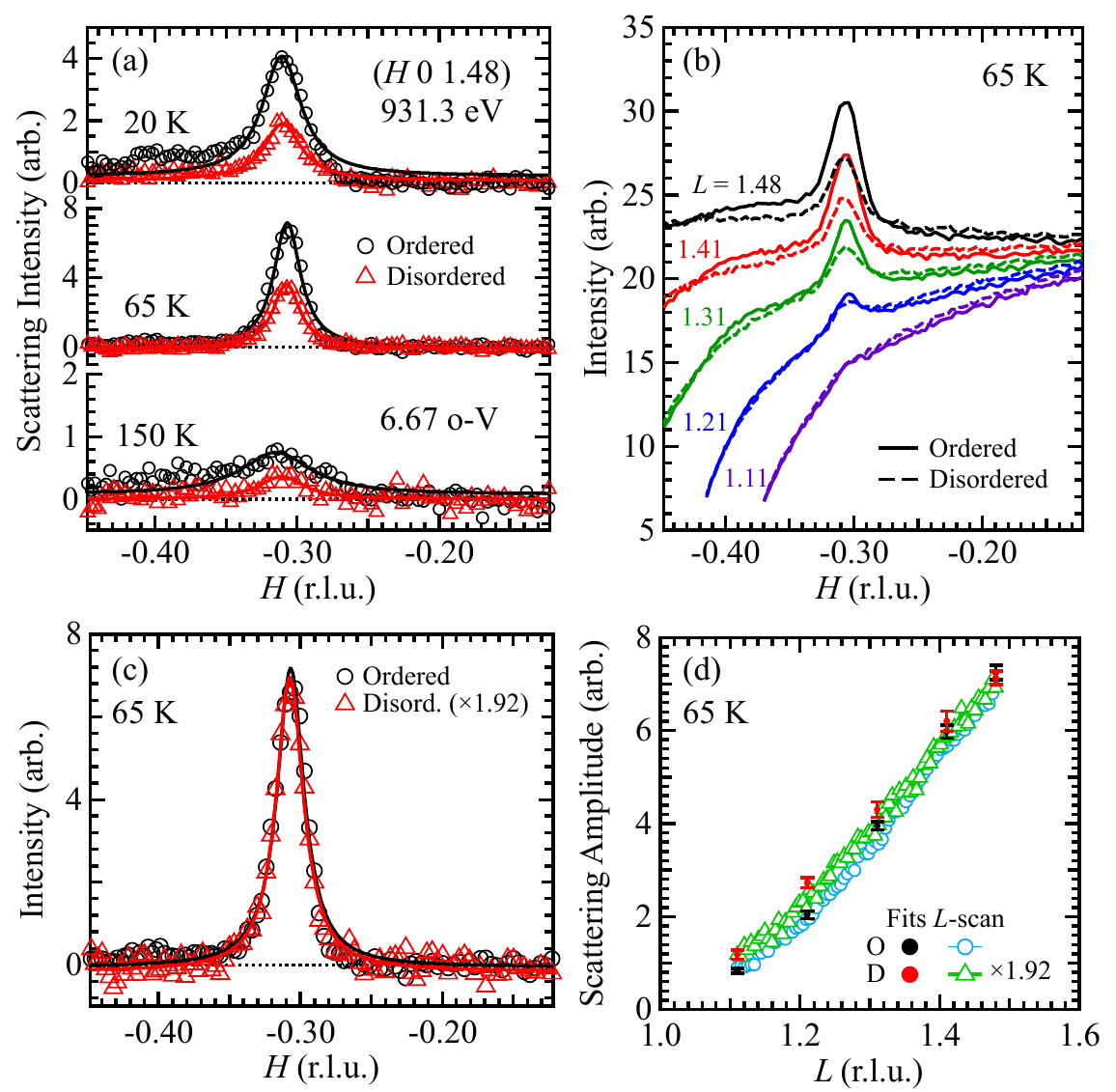}}
\caption{CDW scattering intensity in o-V ordered YBCO before and after quench cooling.  (a) Background subtracted scattering intensity at select temperatures. Solid lines are Lorentzian fits. (b) ($H$ 0 $L$) scans at various $L$ values. (c) and (d) The $H$ and $L$ dependence of the CDW scattering intensity at 65 K scaled to match peak intensities. The peak width and position (c) and $L$ dependence (d) are the same in the oxygen ordered and disordered states. }
\label{fig2}
\end{figure}

\begin{figure}
\centering
\resizebox{\columnwidth}{!}{\includegraphics{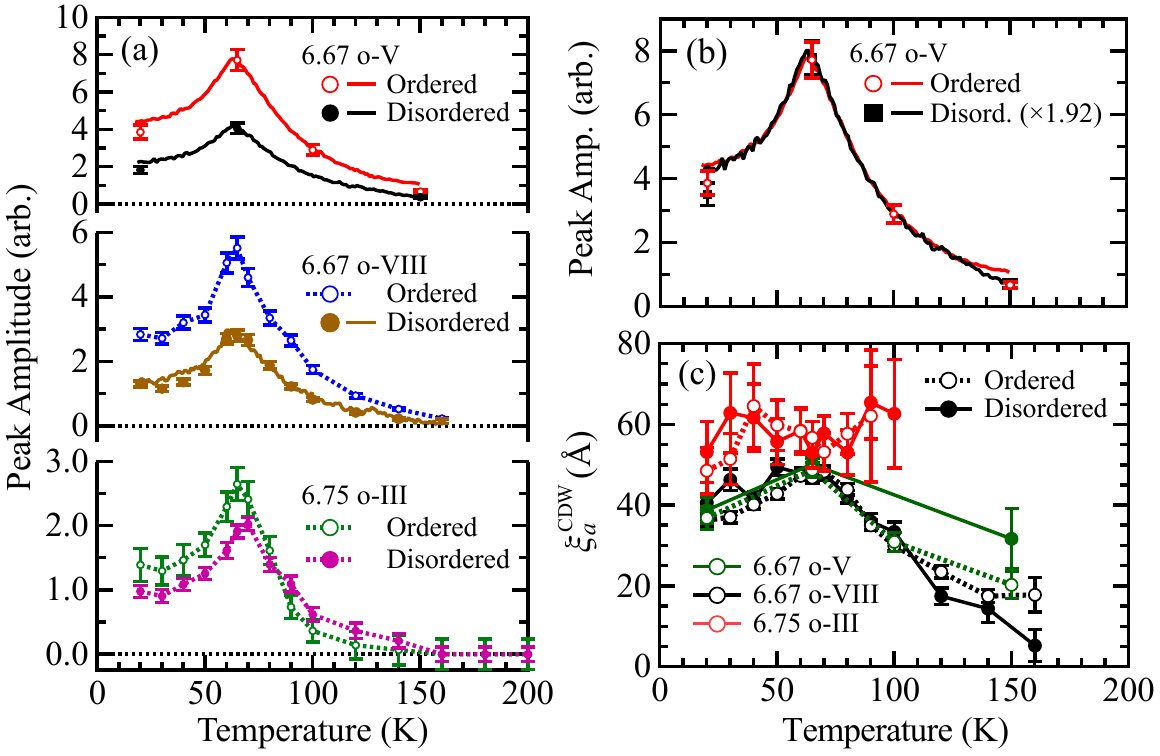}}
\caption{(a) The $T$ dependence of the CDW peak amplitude in oxygen ordered and disordered states. Solid lines (o-V and o-VIII disordered) are the scattering amplitude measured upon cooling and markers are Lorentzian fit amplitudes. Dotted lines connect markers in the cases where scattering amplitude was not measured upon cooling. (b) The $T$ dependence of the CDW peak amplitude in the o-V sample scaled to equal intensity at 65 K. The $T$ dependence is unchanged by disordering the chains. (c) The $T$ dependence of $\xi^\text{CDW}_a$ of the CDW peak in the ordered and disordered states.}
\label{fig3}
\end{figure}

In Fig.~\ref{fig1}(a), RSXS measurements of the CDW peak are shown at 65 K, at $L\simeq 1.48$ and at a photon energy of 931.3 eV, corresponding roughly to the maximum in CDW scattering intensity at the Cu $L_3$ edge.\cite{Ghiringhelli12,Achkar13} The fluorescence background and scattering contributions from nearby oxygen order peaks (off resonance but still visible) were subtracted using a scan at high $T$, as shown in Fig.~\ref{fig2}(a) for the o-V YBCO$_{6.67}$ sample. This procedure was susceptible to a larger error for the ordered o-III YBCO$_{6.75}$ sample due to the overlap of the o-III superstructure reflection ($H\!=\!-0.33$) with the CDW peak ($H\!=\!-0.31$). The correlation lengths [$\xi_a\!=\!a/\left(2\pi\text{HWHM}_H\right)$], $T$ dependence (see Fig.~\ref{fig3}) and incommensurability, as determined by Lorentzian peak fits to the background subtracted data, all vary somewhat weakly with doping and are consistent with previous and more recent work.\cite{Blackburn13,BlancoCanosa14,Huecker14} The measured peak widths were not appreciably influenced by the detector resolution.\footnote{The estimated resolution in the $H$ direction at  $Q\!=\!(-0.31\ 0\  1.48)$ is $\Delta H \leq 0.019$ r.l.u.} Note that in addition to determining the $T$ dependence with fitting of $Q$-scans, we have also monitored the detector count rate while the sample cooled below $T_\text{CDW}$, as shown in the solid lines of Fig.~\ref{fig3}(a). Here we have subtracted a constant value for the fluorescence background, which from measurements at (-0.25 0 1.48) (away from the CDW peak) are found to be weakly temperature dependent over the temperature range of interest.  Accordingly, we find the shape of the cooling curves to be consistent with the peak fitting results.

After measurement in the ortho ordered state, the ortho phase was disordered by heating to $\sim$100 $^\circ$C (above the oxygen ordering temperature). Despite oxygen atoms being mobile at this temperature, the kinetics of oxygen incorporation at the surface are very slow, so no change in sample stoichiometry is expected.\cite{Bobowski10} During this process, the pressure in the chamber was maintained below $6\times10^{-9}$ Torr, ensuring a clean sample environment and no surface contamination. The samples were maintained at 100 $^\circ$C long enough for the oxygen order superstructure reflection to disappear ($\sim$30 to 60 min) and cooled back down to room temperature in $\sim$7 min (dictated by the maximum cooling rate of the instrument) to quench in the oxygen disorder. The degree of oxygen order is characterized by the intensity and width of the ortho order superstructure peaks along the $a$-axis, shown in Fig.~\ref{fig1}(b). The correlation length along the $b$-axis, which is known to be larger,\cite{Chang12} was not measured here. In the ortho VIII ordered YBCO, the o-VIII ordering peak is replaced by weak o-V order at $H\!=\!-0.4$ upon quenching.  Similarly, the quenching procedure destroys the o-V phase in the ortho V crystal and nearly eliminates the o-III phase of the ortho III crystal.

We find that disordering the chains results in a decrease in CDW peak intensity [Fig.~\ref{fig1}(a)], but essentially no change in the CDW peak incommensurability ($Q$ position) [Fig.~\ref{fig2}(c)], temperature dependence [Fig.~\ref{fig3}(a)] or correlation length (either in-plane, $\xi^\text{CDW}_a$ [Fig.~\ref{fig3}(c)], or out-of-plane, $\xi^\text{CDW}_c$ [Fig.~\ref{fig2}(d)]). This is most clearly seen in the $x\!=\!0.667$ o-V sample, where scaling the background-subtracted CDW scattering intensity after quenching by a constant factor of $\sim$1.9 provides an excellent match to the $H$, $L$ [Fig.~\ref{fig2}(c) and \ref{fig2}(d)] and $T$ dependence [Fig.~\ref{fig3}(b)] of the CDW peak in the original oxygen ordered state. Since no change in the $L$ dependence of the peak is observed [Fig.~\ref{fig2}(b) and \ref{fig2}(d)], we find that oxgyen disorder has a negligible impact on $\xi^\text{CDW}_c$.\footnote{The data suggests that $\xi^\text{CDW}_c<c$ in both cases, consistent with the hard x-ray results in Ref.~\cite{Chang12}, but we cannot determine a reliable value given the limited $Q$-space.} 

\begin{figure}
\centering
\resizebox{\columnwidth}{!}{\includegraphics{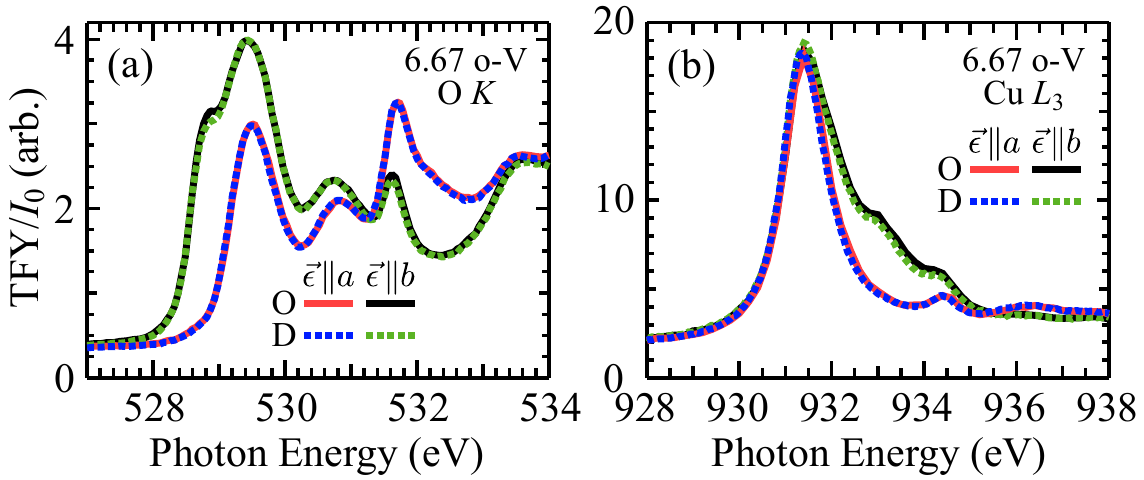}}
\caption{(a) O $K$ and (b) Cu $L_3$ edge XAS of the o-V sample in the oxygen ordered (O) and disordered (D) states with incident photon polarization, $\vec{\epsilon}$, parallel to the $a$ and $b$ crystal axes. Nominally, no change in orbital occupation in the CuO$_2$ planes is seen.}
\label{fig4}
\end{figure}

The disorder-independent incommensurability, $\xi^\text{CDW}$ and $T$ dependence argue against a strong role of the chain order periodicity in stabilizing the CDW order, consistent with the previous observations of distinct $Q$, energy and temperature dependence for the CDW and chain superstructure peaks.\cite{Achkar12}  In addition, the change in CDW intensity does not appear to be associated with a change in the hole doping in the CuO$_2$ planes upon disordering the chains. A change in hole doping might be expected since disordering the oxygen atoms can reduce the chain length -- affecting the charge transfer to the CuO$_2$ planes by reducing the number of Cu atoms in the full chains that are coordinated by 4 oxygen atoms (2 apical and 2 in the chain layer) and increasing the number of Cu that are coordinated by only 3 oxygen atoms. However, XAS measurements (consistent with Refs.~\cite{Nucker95,Hawthorn11b}) before and after the quenching procedure, shown in Fig.~\ref{fig4}, indicate that the hole doping change induced by disordering the chains is negligible, at least in the o-V sample. Moreover, under the premise that the CDW peak is most intense around $p\!=\!1/8$,\cite{Ghiringhelli12} underdoping the $x\!=\!0.75$ sample would presumably enhance the CDW order, whereas the measured effect is a modest decrease in intensity by a factor of $\sim\!1.3$.  

Consequently, rather than a change of hole doping in the CuO$_2$ planes, the decrease in CDW scattering intensity likely results from a change in the defect structure of the materials, specifically how the O disorder in the chain layer influences the CuO$_2$ planes.  These defects include point-like defects due to the ends of finite length chainlets and also domain walls caused by phase slips in the chain ordering pattern. In the O ordered phase, the short correlation length of the chain order implies an already large density of both types of defects. Quenching the samples into an O disordered phase decreases the average length of the chainlets, increasing the density of weak, random point-like defects. However, since the chain order is originally short-range, quenching the samples into an O disordered phase can also decrease rather than increase the domain wall density  (a fully-disordered, random O distribution in the chain layer would have a single o-I domain with no domain walls). Somewhat counter-intuitively then, depending on the interplay of increasing point defect density and decreasing domain wall density, disordering the chains may result in more or less disorder in the CuO$_2$ planes.  In short, although the oxygen atoms in the chain go from being ortho ordered to disordered after quench cooling, it is not yet clear whether the resultant decrease in CDW intensity is due to an increase or decrease in defect density in the CuO$_2$ planes. Future work to determine the proportion of these defects is needed to clarify this point. We also note that the variation in disorder here is unlikely to correspond to a variation in the strength of point-like defects, making it unclear how our measurements correspond to existing theoretical models of disorder effects on CDW order.\cite{Robertson06,DelMaestro06} 

We note that the measured $\xi^\text{CDW}$ (at $\sim T_c$) both here and in a wider range of dopings is only weakly dependent on the excess oxygen content.\cite{BlancoCanosa14,Huecker14}  Since the level of oxygen disorder varies considerably in the investigated range of samples, a substantial doping dependence to $\xi^\text{CDW}$ would be expected if oxygen disorder were setting the length scale of the CDW order. 

The effect of O disorder shown here should be compared with the effect of Zn-impurities on the CDW order in YBCO.\cite{Blanco-Canosa13} Blanco-Canosa \textit{et al.} \cite{Blanco-Canosa13} found that Zn doping decreases the CDW peak intensity, like our O disorder measurements.  But, unlike O disorder, Zn-impurities decrease the correlation length and significantly change the $T$ dependence of the CDW order.  This was argued to be consistent with spatially inhomogeneous CDW order, with CDW order suppressed in regions around the Zn impurities, where incommensurate SDW order is enhanced.  However, it is not clear that a similar description can be applied to O disorder since Zn doping introduces a spinless impurity in the CuO$_2$ planes that is a much stronger defect than O defects in the chain layer and results in strong pair breaking scattering, suppression of superconductivity, a slowing of spin fluctuations and of  static, incommensurate SDW order.  Moreover, the lack of dependence on $\xi^\text{CDW}$ and $T$ dependence for our O disorder measurements are difficult to reconcile with a simple inhomogeneity model, where the volume fraction of CDW order is decreased when the O in the chain layer is disordered.  Since $\xi^\text{CDW}$ is not impacted by the O disorder, this scenario would require a domain size larger than the CDW correlation length.  Given that the defect density is high in the O disordered state, and that the correlation length in the ortho ordered states is in fact less than that of the CDW order [the O order is short range, see Fig.~\ref{fig1}(b)], it is unclear how CDW domains would form around O defects. 

It is also difficult to reconcile the independence on oxygen disorder of the CDW $T$ dependence and correlation length with pinning of CDW order that would be fluctuating in the absence of disorder, since the temperature evolution of the fluctuations should depend on the level of disorder.\cite{Nie13}  Rather, these measurements suggest that the short range nature of the CDW order ($\xi^\text{CDW} < 60$\ \AA) could have an intrinsic origin related to the competition between CDW order and SC rather than being a result of sample disorder. For instance, it has been shown that the gradual, concave $T$ dependence of the CDW order shown in Fig.~\ref{fig3} can result from angular fluctuations of a multi-dimensional order parameter comprised of biaxial CDW order and superconductivity.\cite{Hayward13}  Although the impact of disorder on this model has not been investigated, it is plausible that disorder can affect the CDW peak amplitude without having a strong effect on the angular fluctuations that govern the CDW $T$ dependence and possibly the correlation length.  

Although a detailed explanation for the O disorder dependence is yet not evident, in the absence of a change in CDW volume fraction or pinning, it is reasonable to consider the reduction of the CDW peak intensity as being associated with a reduction in the magnitude of the CDW order parameter.  As previously demonstrated, the CDW peak intensity at the Cu $L$ edge is associated with a spatial modulation in the Cu 2$p$ to 3$d$ transition energies $\Delta E$, which for small modulations scales as $\Delta E^2$.\cite{Achkar12,Achkar13}  Accordingly, the magnitude of these energy modulations, which are presumably proportional to the CDW order parameter, could be affected by quenched oxygen disorder (and full details of how quench cooling alters the defect structure). In this context, we note that the CDW order is more strongly reduced by disorder in the YBCO$_{6.67}$ samples than it is in the YBCO$_{6.75}$ sample.  Although, we should caution that the level of disorder is not well calibrated between samples, it is curious that the larger change in the magnitude of the CDW order occurs in samples where the CDW peaks are most intense.  Naively one may have expected CDW order to be more susceptible to disorder at doping levels where the order is weaker.  

Finally, we address how these results on YBCO can be understood in the broader context of CDW order in the cuprates. We showed here that the degree of oxygen disorder in YBCO does not determine $\xi^\text{CDW}$, suggesting a more intrinsic origin for the short range character of the CDW order. Although YBCO represents a case of relatively weak disorder, the implication that disorder is not the dominant factor in determining $\xi^\text{CDW}$ may be applicable to other cuprates. This may explain why across the various cuprate families, where the type of defects and their importance varies considerably (e.g. cation substition, oxygen vacancies, lattice distortions),\cite{Eisaki04} $\xi^\text{CDW}$ is of the same order of magnitude (e.g. 20--30 \AA  ~in Bi-2201,\cite{Comin14a} $\simeq 20$ \AA  ~in Hg-1201,\cite{Tabis14} and 50--175 \AA  ~in LBCO\cite{Hucker11}). Rather than an emphasis on disorder, an understanding of the competition between CDW order and superconductivity, and how this competition is influenced by the electronic and crystalline structure (e.g. Fermi surface topology, interlayer coupling), may clarify how the CDW order differs in the various cuprate families. 

\begin{acknowledgements}
We thank S. Sachdev, G. A. Sawatzky, A. Burkov, M. Le Tacon and B. Keimer for useful discussions. This work was supported by the Canada Foundation for Innovation, the Canadian Institute for Advanced Research and the Natural Sciences and Engineering Research Council of Canada.  Research described in this paper was performed at the Canadian Light Source, which is funded by the Canada Foundation for Innovation, the Natural Sciences and Engineering Research Council of Canada, the National Research Council Canada, the Canadian Institutes of Health Research, the Government of Saskatchewan, Western Economic Diversification Canada, and the University of Saskatchewan. A.J.A., X.M. and C.M. acknowledge the receipt of support from the CLS Graduate Student Travel Support Program.
\end{acknowledgements}

\bibliographystyle{rmp}
\bibliography{RSXSbibYBCO_Odisorder2}

\end{document}